\documentclass[a4paper,12pt]{article}
\usepackage[utf8x]{inputenc}
\usepackage{graphicx}
\usepackage[left=2.5cm,top=2.5cm,right=2.5cm,left=2.5cm]{geometry}
\usepackage{hyperref}
\usepackage{color}
\usepackage{amsmath}
\usepackage{amsthm}
\usepackage{amsfonts}
\usepackage{mathrsfs}
\usepackage{multicol}
\usepackage{bm}
\usepackage[noblocks,affil-sl]{authblk}
\usepackage{footnote}
\usepackage{float}
\usepackage[small,bf,hang]{caption}
\usepackage{rotating}
\usepackage{hvfloat}
\usepackage{graphics}
\usepackage{subfig}
\usepackage[all]{xy}
\usepackage{subfig}
\usepackage{longtable}
\usepackage{latexsym}

\title{\textbf{Occupational mobility network of the Romanian higher education graduates}}

\author[1,2]{Eliza-Olivia Lungu}
\author[1]{Ana-Maria Zamfir}
\author[1]{Eva Militaru} 
\author[1]{Cristina Mocanu}
\affil[1]{\small{National Scientific Research Institute for Labour and Social Protection, 6-8 Povernei St., District 1, 010643 Bucharest, Romania}}
\affil[2]{\small{The Bucharest Academy of Economic Studies, 6, Romana Square, District 1, 010374 Bucharest, Romania }}

\begin{document}

\maketitle

\begin{abstract}
Although there is a rich literature on the rate of occupational mobility, there are important gaps in understanding
patterns of movement among occupations. We employ a network based approach to explore occupational mobility of the 
Romanian university graduates in the first years after graduation ($2003$ - $2008$). We use survey data on their career 
mobility to build an empirical occupational mobility network ($OMN$) that covers all their job movements in the considered period.
We construct the network as directed and weighted. The nodes are represented by the occupations (post coded at 3 digits according to 
ISCO-$88$) and the links are weighted with the number of persons switching from one occupation to another. This representation of data
permits us to use the novel statistical techniques developed in the framework of weighted directed networks in order to
extract a set of stylized facts that highlight patterns of occupational mobility: centrality, network motifs. 
\end{abstract}

\section{Introduction} \label{intro}
Occupational mobility represents an important feature of the labour markets. While there is a rich literature on the structure 
and rate of occupational mobility, 
there are important gaps in understanding the career trajectories. Such a challenge should rely on understanding individual 
histories of transitions among 
occupations into coherent sequences, if such patterns exist. This paper explores the early career of higher education 
graduates during the first years after graduation in order to understand 
their occupational mobility. For understanding paths of occupational 
mobility, we unify elements 
from job mobility and human capital theories. By exploiting a unique dataset on working histories of higher 
education graduates from Romania, 
we provide novel evidence on the fact that individuals move to similar occupations and that the entrance 
occupation influence their subsequent career.

During the last two decades, participation in higher education increased significantly in Romania. Labour Force Survey (LFS) data 
show that the share of higher education 
graduates in active population increased from $7.9\%$ in $1997$ to $14.3\%$ in $2008$. Romanian economy has witnessed dramatic 
changes associated in the $'90$ with an in-depth 
transition from plan to market and then with high rates of economic growth between $2000$ and $2008$. 
Job creation during the last decade was mainly 
concentrated in sectors and occupations placed at the bottom of the added value chain. 
Thus, it is obvious that the demand side and the supply side 
of the Romanian labor market have evolved in an uncorrelated manner. As a result, high shares of 
higher education graduates enter
the labour market in mismatched jobs. From this point of view, occupational mobility in the first years of their 
career can offer them the chance of reaching 
jobs more appropriate for their level of education.   

We investigate patterns of occupational mobility by employing a network based approach in 
which nodes are represented by occupations and links by the flows of individuals moving from one occupational category 
to another. Such an approach helps us to better visualize paths of mobility and calculate network 
indicators in order to understand models of connectivity between different occupations.
We consider that occupations are related to each other via transferable skills which can be sector-specific or 
not. 

The paper is structured as follows: section 2 revises the main theories in occupational mobility and youth career, 
section 3 describes the database that we are using, section 4 presents the methodology and the results,
while section 5 concludes and outlines directions for future study.

\section{Occupational mobility in the early stage of the career post graduation} 
The classic model of turnover relies on the fact that workers look for a better match to the type or work 
performed and to the employer \cite{Jovanovic}; \cite{Neal}
Empirical evidences suggest that matching takes place at occupational level as information obtained by individuals 
working in a job is used to predict the quality of the match 
at other jobs within the same occupation. Thus, those working their first job are more likely to leave the current 
job than those working their second job in the 
same occupation \cite{McCall}. 

Separately from the job mobility theory, an approach in which occupations are related to each other was developed. 
Becker's human capital theory 
conceptualizes the existence of general and firm-specific human capital. In subsequent theoretical developments, 
human capital includes industry-specific skills \cite{Neal}, 
while investigation of wage formation showed that there are occupation-specific skills which are transferable 
across employers. This means that when a worker switches the 
employer or the sector, he or she loses less human capital than when changing occupation \cite{Kambourov}. 
Higher loss of human capital represents higher costs 
for mobile workers. However, some occupations are linked to each other due to the transferability of skills. 
Such occupations in which skills and experience can 
be partially or fully transferred from one occupation to another form career paths \cite{Weiss}; \cite{Shaw}. 
\cite{gathmann200701}
demonstrate that some occupations are 
related to each other by exploiting data on task requirements. An additional theoretical development 
is the concept of career ladder which means that some occupations form a 
path of advancement for workers gaining skills and experience \cite{Sommers}. Workers move up 
on the ladder as they gain skills and information 
on their productivity \cite{Sicherman}. However, patterns of occupational mobility cannot be explained 
only by advancements of individuals on a 
career ladder \cite{gathmann200701}. On the other hand, a career line or job trajectory 
was defined as a work history that is similar for a part 
of labor force. 
Group disparities in patterns of occupational transitions should portray different career lines. 
So, a career line represents a collection of jobs which is 
characterized by a 
high probability of movement from one occupation to another for certain groups of individuals \cite{Spilerman}. 

Job mobility represents an important characteristic of the first years of employment as rate 
of job changing declines with age and experience of the 
workers \cite{Topel}; \cite{Davia}. From the perspective of human capital theory, youth inexperience lower 
costs of occupational mobility due to the fact 
that they have invested less in their occupational-specific skills. In the literature, this period is called 
the shopping and thrashing stage and is characterized by exploration and 
testing of the market.   

\cite{Isoda} analyis the occupational mobility as the geographical mobility and generates a 
job map using the occupational change data from the British Quarterly Labour Force Surveys 2001-2005. The map
has on x-axis the skill level and on y-axis the skill contect of the occupations, so similar ones are pleace nearby 
and disimilar occupations far away. This type of mapping is intended to help the job seekers in the process of building their carear.
The main conclusions of the study are: most of the entrances on the labour market are in elementary or low skilled jobs; main 
direction of the occupation change is in career ugrating and the retirement is present equally among occupations.

By employing a network-based approach, we complement his work producing new evidences for the 
usefullnes of alternative methods in systematic understanding of the relationships 
among occupational categories. 

\section{Data}
The database employed for this study is based on a national survey carried out between November, 2008 - January, 2009. 
The survey was designed to
investigate the labor market entry process and early career of the Romanian university graduates in the first 5 years after 
graduation.The data were retrospectively recorded and they 
cover both personal characteristics and information regarding the jobs 
accessed by the subjects from graduation until the moment of 
investigation: professional status, 
job title, occupation, industry type,
type of contract, how the job was 
found e.g. In total there were investigated
$2194$ university graduates, the sample being stratified on their 
educational profile according to the 
structure provided by Romania's National Institute of Statistics. 

Originally, the occupations were recorded as string variables, but
for this analysis we post-coded them according to the International Standard Classification of Occupations
ISCO-88 at 3 digits, in order to keep under control the
measurement errors in occupational codes. ISCO-88 specify a system for classifying and aggregating
occupations based on the degree of 
similarity of their constituent tasks and duties. It is the result of a series of lengthy
and detailed investigations in twelve EU countries, combining the knowledge of experts in occupational classification in each
country with the practical considerations for coding occupational information collected by census and survey techniques.
This classification organises the occupations in an hierarchical framework at the lowest level beging the job,
defined as a set of tasks and duties designed to be executed by one person. 
Although each job may be distinct in term of outputs required, they are sufficiently similar in terms of the abilities 
needed as inputs in order to be regarded as a single occupational unit for statistical purposes. 
At 3 digits there are in total $130$ occupations. 

At the moment of survey investigation, $93.8\%$ of university graduates interviewed were employed. 
This high insertion rate of the university graduates 
is explained mainly by the moment when the survey was carried out, during that period the Romanian economy was 
still growing and the labor force demand was at 
the highest level witnessed in the last $20$ years. Thus, we have to underline that our
 analyses is carried out in a specific labor market context with 
low unemployment rate, high speed of filling vacancies and high job creation. 
On the labour market, three types of occupational mobility may appear:
newly entrants, job change across occupations
and labour market exits, for exemple due to retirement. This article is focused strictly
on the second type of occupational mobility and we considere their movements as voluntary.
Investigating the employment patterns of 
university graduates in the first years post graduation, we noticed two things: they were practically 
rapidly ``absorbed'' by 
the labour market and they encounter a significant job mobility ($30\%$ changed the job at least ones and around $10\%$ changed
their occupation). For this study we follow all the job changes even if they result or not
in an occupational change. 

\section{Occupational mobility network}
In the recent period, a large body of literatures employing a network-based approach has been emerging in the study 
of socio-economic systems \cite{Granovetter}; \cite{Freeman}; \cite{Barabasi1}; \cite{Caldarelli};
\cite{Fagiolo2}.
Within this approach, the socio-economic systems such as markets, industries or even the global 
economy are viewed as networked structures. Between the main studied topics are:
international trade \cite{Li}; \cite{Reichardt}, social relations structures 
\cite{Jackson1}, 
internet, peer-to-peer networks,
electronic circuits, neural networks, metabolism and protein interactions etc. The network-based approach is
employed in such a large diversity of subjects because it 
permits assessment of several underlying indicators (maximum distance between pairs of nodes, clustering 
tendency, distribution of degrees of the nodes, motifs, centrality measures etc.) which characterize the investigated structures.

We employ a network approach to analyse the statistical properties of the occupational mobility network ($OMN$) 
of the Romanian university graduates in a five years 
time frame ($2003$ - $2008$). We use the records about their career mobility to built an occupational mobility 
networks ($OMN$) that covers all their job shifts in the considered period.
We generate the network by overlaying the transition matrix between jobs, without
taking into account the unemployment spells between jobs. We construct the network as
directed and weighted. It is important in this case to consider both the magnitute and the direction of the
flow of workers between occupation in order to have a complete view. The nodes represent 
the occupations postcoded at 3 digits according to ISCO-88 and 
the links are weighted with the number 
of persons moving from one occupation to another. We also counted the people that are changing their jobs 
in the same occupations as self-loops. An occupation change means a modification in the occupational code at 3 digits at the
trasition from one job
to another.
\begin{figure}[h]
 \centering
 \includegraphics[scale=0.13]{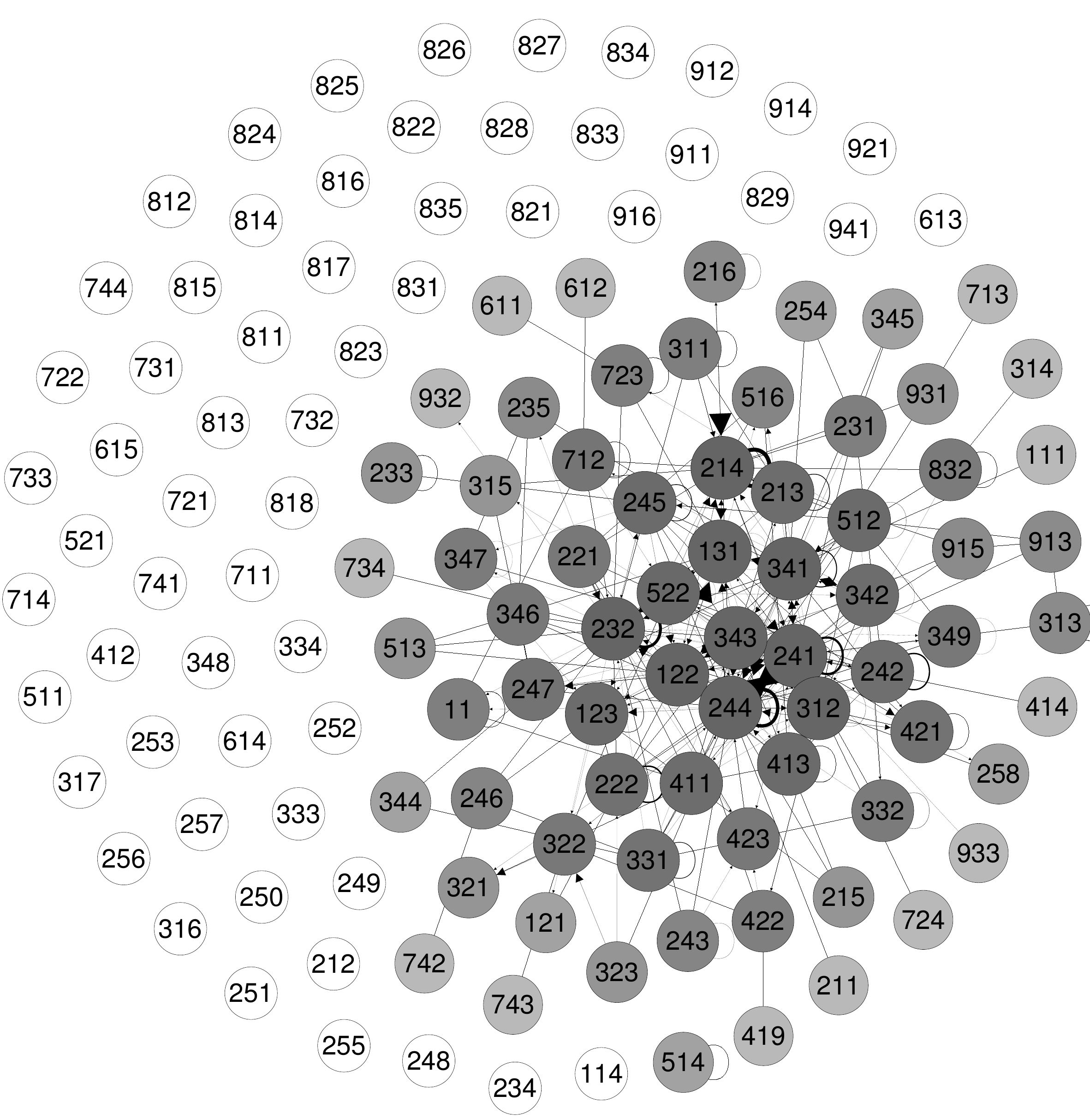}
 \caption{Visualization of the empirical occupational mobility network (OMN).  
The thickness of the links is proportional to their weights and the color of the nodes reflects their total strenght (white - 
isolated node ($48\%$ of all the nodes), black - highly connected node). The numbers in vertexes represent the occupational codes
according to ISCO-88.}
 \label{fig: reteatotal}
\end{figure}
Knowledge of such topological properties is esential, in order to have a global perspective over the occupations set and how they
connect.
Also, while interpreting the results is crucial to keep in mind the specificity of the country and the 
considered time frame. 
We represent $OMN$ as a graph $G(N)$ with $N=130$ nodes, representing the total number of occupations at 3 digits according 
to ISCO-8 and $n=322$
edges. The
real-valued $NxN$ matrix $A={a_{ij}}$ has positive elements if there is a flow of persons between occupations (nodes).
The graduates that are changing their job in the same occupation are represented in $OMN$ as self-loops \ref{fig: reteatotal}.
The network density is $0.016$ and it describes 
the general level of linkage among the nodes in the graph. The average degree is $4.15$ and 
while the average clustering coefficient is $0.09$.

\subsection{Node Centrality}
Node centrality is a key issue of the social networks. The relevance of this measure
emerges from the fact that job mobility defines a certain degree of dependency 
of an occupation to another. In this case the 
vertex (occupation) centrality denotes the likelihood of a given occupation to appear along a randomnly selected 
mobility flow within the $OMN$. The higher is their likelihood the more
influencial is the occupation in the network. So shocks affecting central occupation are more likeliy 
to be transmited in the whole network, to all the occupations.
\begin{figure}[h!]
  \centering
  \subfloat[]{\label{fig:fig1}\includegraphics[width=0.5\textwidth]{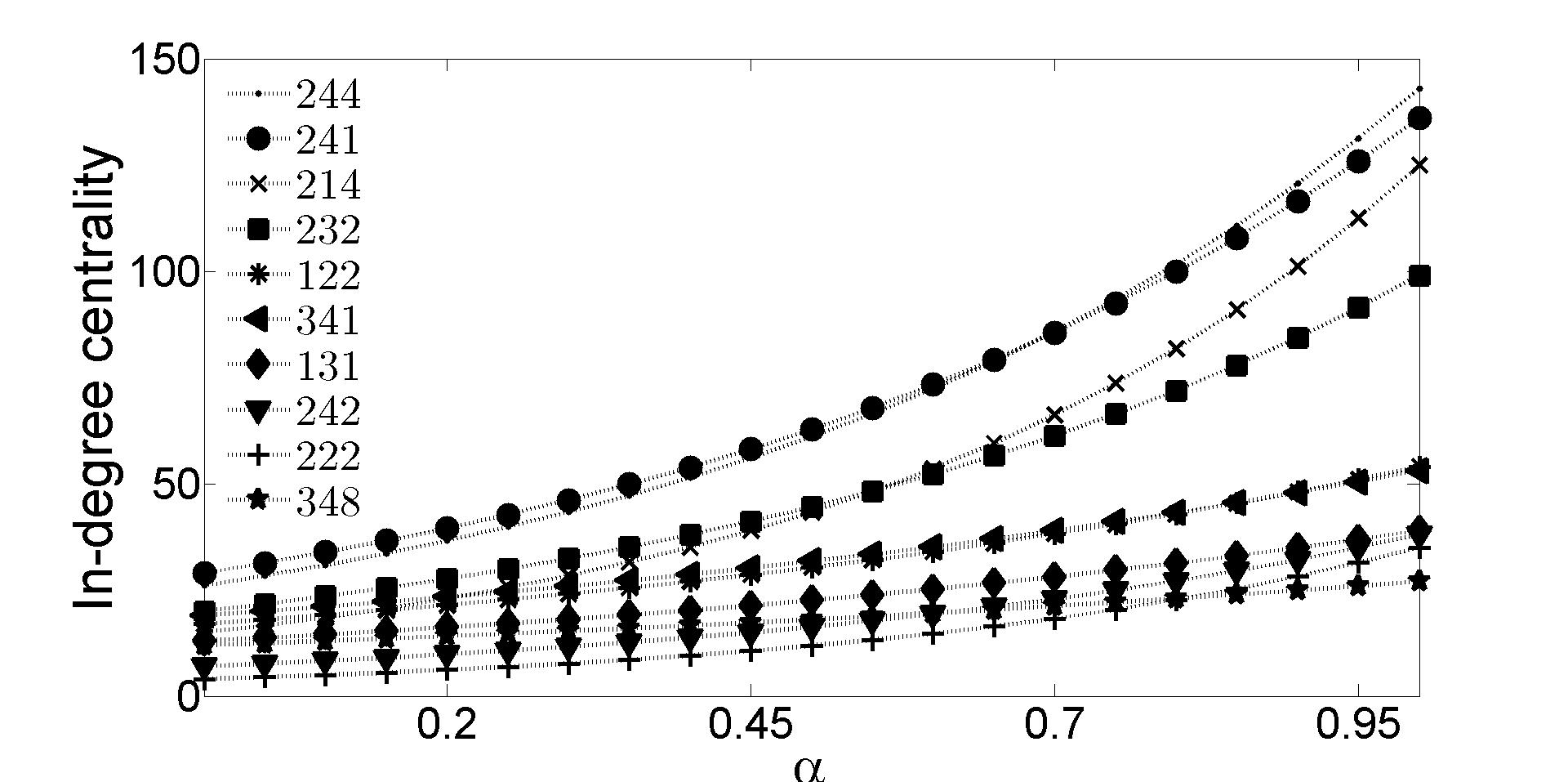}}                
  \subfloat[]{\label{fig:fig2}\includegraphics[width=0.5\textwidth]{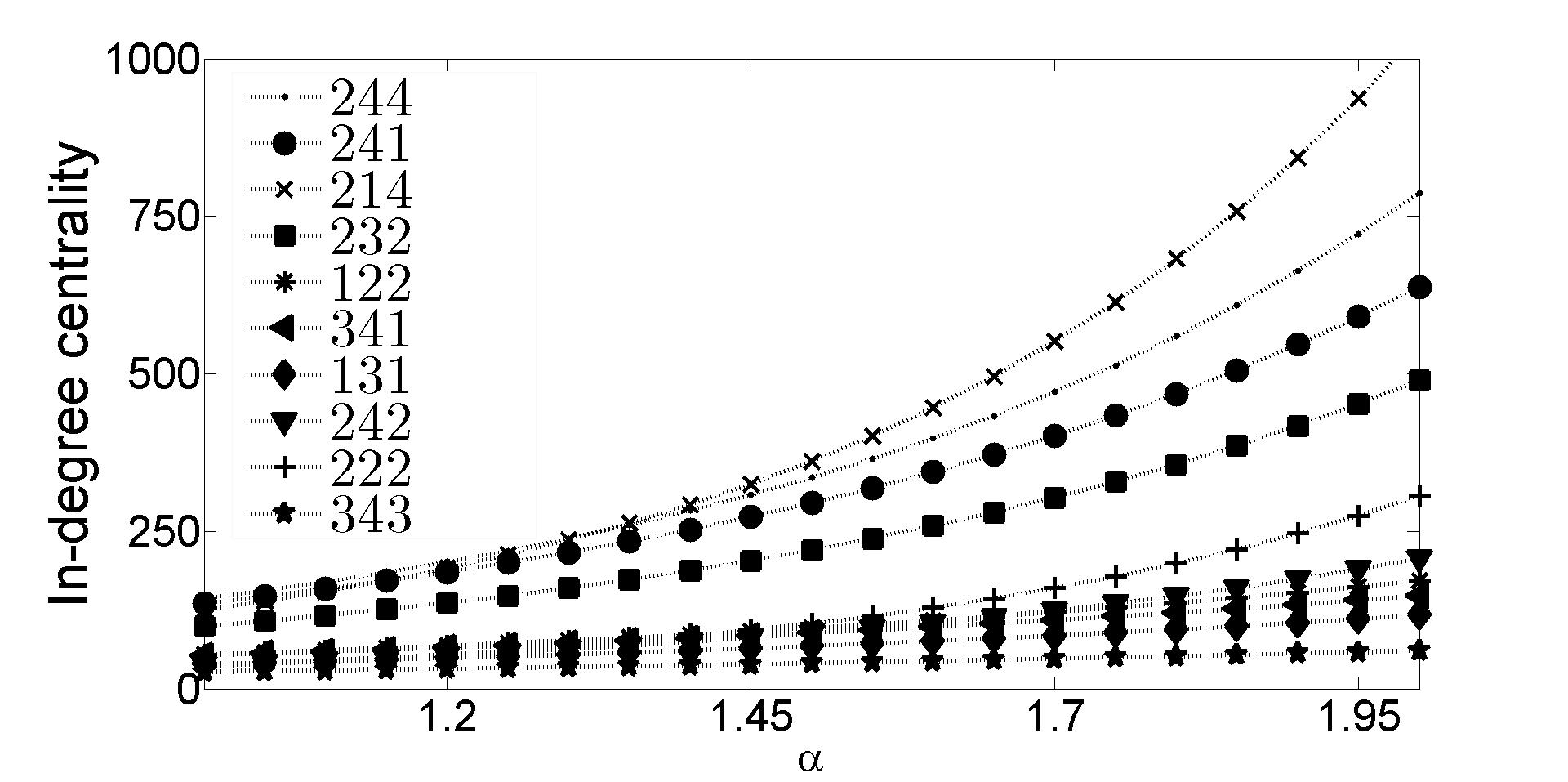}} \\
  \subfloat[]{\label{fig:fig3}\includegraphics[width=0.5\textwidth]{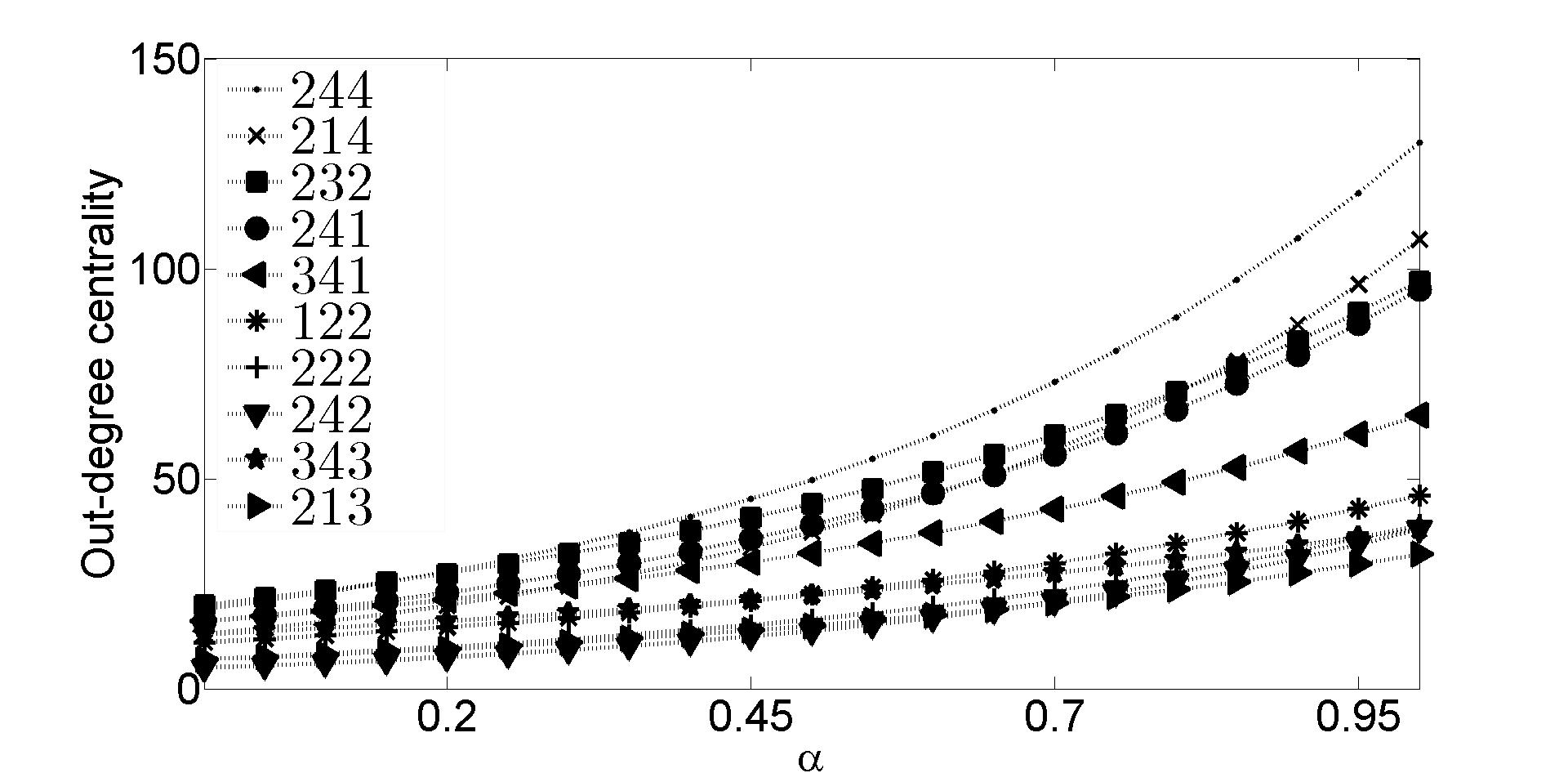}}
  \subfloat[]{\label{fig:fig4}\includegraphics[width=0.5\textwidth]{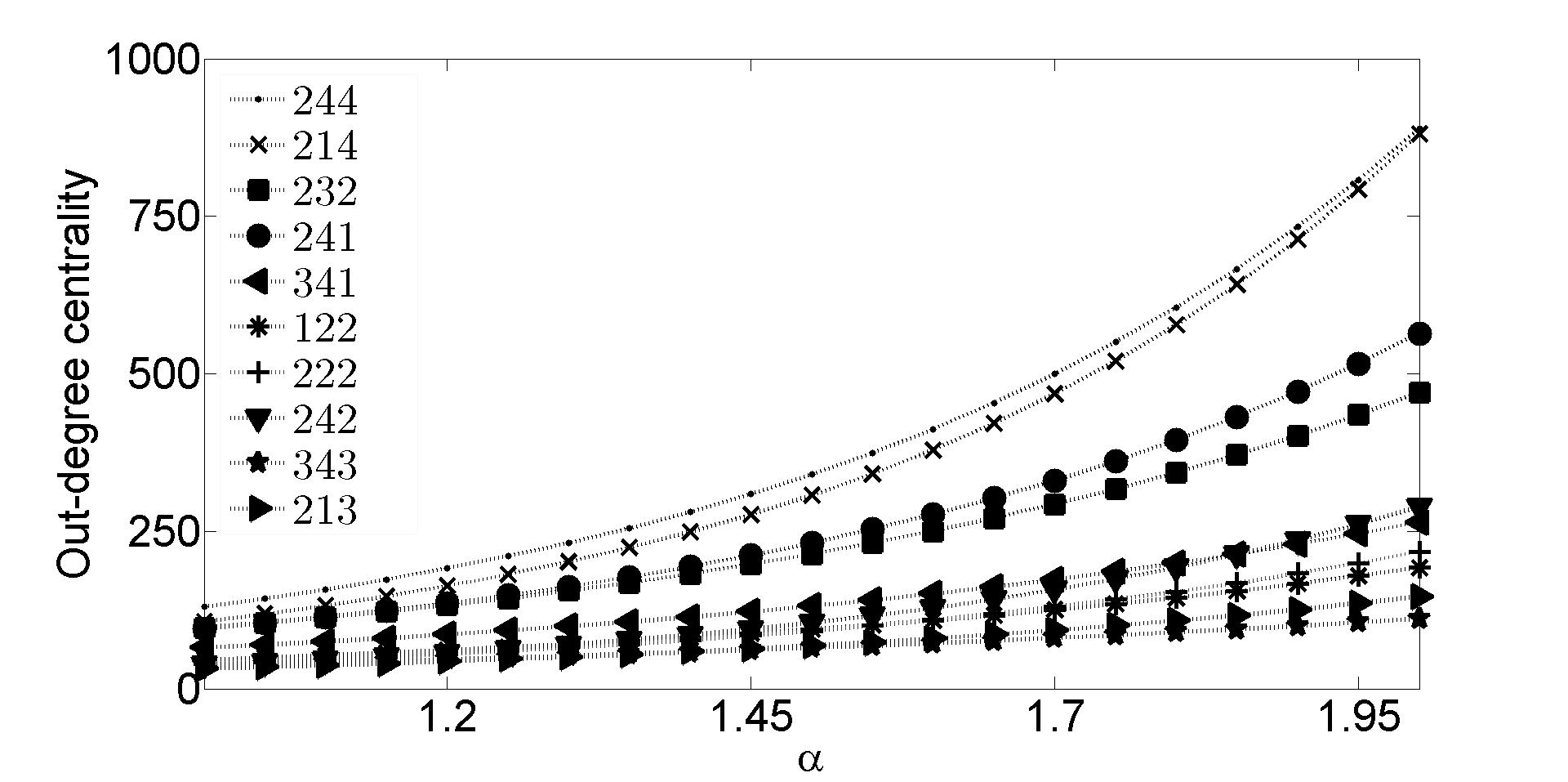}}
  \caption{Degree centrality scores when different values 
of $\alpha$ are used: $\alpha \in [0,1]$ $(a,c)$ and $\alpha \in [1,2]$
$(b,d)$}
  \label{fig:degree}
\end{figure}
We consider three measures to quantify the centrality of a node: degree centrality, closeness and betwenness. 

The degree
centrality has been introduced by \cite{Opsahl2010245} and it combines both degree and
strength with a positive tuning parameter $\alpha$ that controls for their relative importance. 
We distingues between in and out degree centrality using (\ref{eq2}) and (\ref{eq3}):

\begin{equation}
\label{eq2}
C_{D_{out}}^{w\alpha}(i)=k_i^{out}*(s_i^{out}/k_i^{out})^{\alpha} 
\end{equation}
\begin{equation}
\label{eq3}
C_{D_{in}}^{w\alpha}(i)=k_i^{in}*(s_i^{in}/k_i^{in})^{\alpha} 
\end{equation} 
where $k_i^{out}$ is the out-degree, $k_i^{in}$ is the in-degree, 
$s_{i}^{out}$ is the out-strenght, $s_{i}^{in}$ is the in-strenght and $\alpha$ a real positive number.

If $\alpha=0$ the in/out node centrality is equal with in/out degree and if $\alpha=1$ the in/out node centrality is equal with
the in/out strenght. When $\alpha \in (0,1)$ more importance is given to the number of ties versus tie weights and
if  $\alpha>1$ vice versa. 

In order to see how sensitive is the node centrality to small variations of the parameter $\alpha$ we
performed a sensitive analysis on the first 10 vertexes according to their in-strenght (\ref{fig:fig1}, \ref{fig:fig2}) and
out-strengh (\ref{fig:fig3}, \ref{fig:fig4}). The considered nodes are involved in $70\%$ of the total in-flow and out-flow of the
empirical network. In the case of in-degree centrality, we notice that when $\alpha \in [0,1]$, the occupations that have
the highest inflow of graduates are social science and related professionals $(244)$, business professionals $(241)$ and after
that are architects, engineers and related professionals $(214)$, secondary education teaching professionals $(232)$ and at
the bottom all the other 6 occupations.
When we vary $\alpha \in [1,2]$ the most central occupation becomes architects, ingineers and related professionals $(214)$
and after that $244$, $241$, $232$ etc. These means that social science and related prefessionals $(244)$ and business
professionals $(241)$ receive university
graduates from a lot of occupations, but with whom the strenghs are rather low, while architecs, engineers and related
professional $(214)$ receive work force from a limited set of occupations with whom they have strong ties. If we look at this
issue from the transferable skills perspective, it means that it's smoother the transfer from other occupations
to social science related ones then to architects, ingineers and related occupations which are more specialized and technology
intensive. 

If we look at the out-degree centrality for $\alpha \in [0,1]$ we see that again $244$, $214$ and $232$ are the occupations from
where a lot of the university graduates are leaving at one point in their early career. In the case of $244$ it could be explained
by being a ``transit'' occupation and for $232$ by the low wages that are typicaly here. This order is preserved also for $\alpha
\in [1,2]$.

Because a lot of higher education graduates are changing their job in the same occupation during their early career we
eliminate the self-loops and perfome the same sensitivy analysis.
In this case we notice a relatively low  in flow of graduates into the occupational group 214 while for the occupations 244, 232 the
high flow (both in and out) of persons is preserved.  This shows us that the group 214 has many characteristics of a profession.
What distingues a profession from other occupational forms is the balace between general and specialized knowledge. Engineers
have long periods of training, a good balance between general and specialized knowledge but they can also take further
responsbilities by becoming managers and executives, in which case they move into occupational group 232
\cite{volti2008introduction}. Also after '90 there
had been a great influx of gradutes in economics and this is visualized in the high dynamics of group 244. 
 
Closeness centrality is defined as the inverse sum of the shortest distance to all other nodes from a specific node. 
\begin{equation}
\label{eq4}
C_C(i)=[\sum_j^N{d(i,j)}]^{-1}
\end{equation}
where $d(i,j)$ is the lenght of shortest path between two nodes.
According to this node centrality measure the first four nodes are: 241 (Business professionals),
244 (Social science and related professionals), 122 (Production and operations managers)
and 131 (Managers of small enterprise). This shows again that managers, social science
and business related occupations are the best connected in the network. 

Betweenness centrality is the average number of short paths between pairs of the nodes that pass between a certain
node.
\begin{equation}
\label{eq5}
C_B(i)=\frac{g_{jk}(i)}{g_{jk}}
\end{equation}
where $g_{ij}$ is the number of shortest paths between two nodes and $g_{jk}(i)$ is the
number of shortest paths that pass node $i$.

In this case the first four occupations are: 244 (Social science and related professionals),
241 (Business professionals), 232 (Secondary education teaching professionals) and
341 (Finance and sales associate professionals).

\subsection{Network motifs}
The concept of network motifs had been introduced by \cite{Milo1} to denote ''patterns interconnections occuring in
complex networks at numbers that are significantly higher than those in randomized networks''. Studies regarding the
occurance of these patterns had been done in several fields: protein interaction network \cite{Wuchty2003}; 
inter-firm network \cite{Ohnishi_JEIC2010}; complex biological, technological and sociological networks \cite{Milo04}.
For this study we are interesed in the identification of the three-nodes connected motifs, Fig. \ref{fig:motifs} shows all the 
possible combinations.

\begin{figure}[h!]
  \centering
  \subfloat[M1]{\label{fig:fig5}
$$
    \xymatrix @=1.1pc{ \bullet  & \bullet  \ar[l]\\
               \bullet \ar[u]  }
$$}              
  \subfloat[M2]{\label{fig:fig6}$$
    \xymatrix @=1.1pc{ \bullet  \\
               \bullet \ar[u]  & \bullet \ar[l]   }$$ }  
  \subfloat[M3]{\label{fig:fig7}$$
    \xymatrix @=1.1pc{ \bullet  \\
               \bullet \ar[u] \ar[r] & \bullet } $$ }
  \subfloat[M4]{\label{fig:fig8} $$
    \xymatrix @=1.1pc{ \bullet \ar[d] & \bullet \ar[l] \\
               \bullet \ar[u]  } $$ }
  \subfloat[M5]{\label{fig:fig9} $$
    \xymatrix @=1.1pc{ \bullet & \bullet \ar[l] \ar[ld] \\
               \bullet \ar[u]   } $$ } 
  \subfloat[M6]{\label{fig:fig10} $$
 \xymatrix @=1.1pc{ \bullet \ar[d] \ar[r]  & \bullet \\
               \bullet \ar[u]  } $$}
  \subfloat[M7]{\label{fig:fig11} $$
    \xymatrix @=1.1pc{ \bullet \ar[r]  & \bullet \ar[ld] \\
               \bullet \ar[u]  } $$ } \\
  \subfloat[M8]{\label{fig:fig12} $$
    \xymatrix @=1.1pc{ \bullet \ar[d]  & \bullet \ar[l] \ar[ld] \\
               \bullet \ar[u]  } $$ } 
  \subfloat[M9]{\label{fig:fig13} $$
    \xymatrix @=1.1pc{ \bullet \ar[r] \ar[d]  & \bullet \ar[l] \\
               \bullet \ar[u]  } $$ } 
  \subfloat[M10]{\label{fig:fig14} $$
    \xymatrix @=1.1pc{ \bullet \ar[d]  & \bullet \ar[l] \\
               \bullet \ar[u] \ar[ur]  } $$ }
  \subfloat[M11]{\label{fig:fig15} $$
    \xymatrix @=1.1pc{ \bullet   & \bullet \ar[l] \ar[dl] \\
               \bullet \ar[u] \ar[ur]  } $$ }
  \subfloat[M12]{\label{fig:fig16} $$
    \xymatrix @=1.1pc{ \bullet \ar[r] \ar[d]  & \bullet \ar[l] \ar[dl] \\
               \bullet \ar[u]  } $$ }
  \subfloat[M13]{\label{fig:fig17} $$
    \xymatrix @=1.1pc{ \bullet \ar[r] \ar[d]  & \bullet \ar[l] \ar[dl]  \\
               \bullet \ar[u] \ar[ur] } $$ } 
  \caption{All types of three-node connected subgraphs (motifs)}
\end{figure}
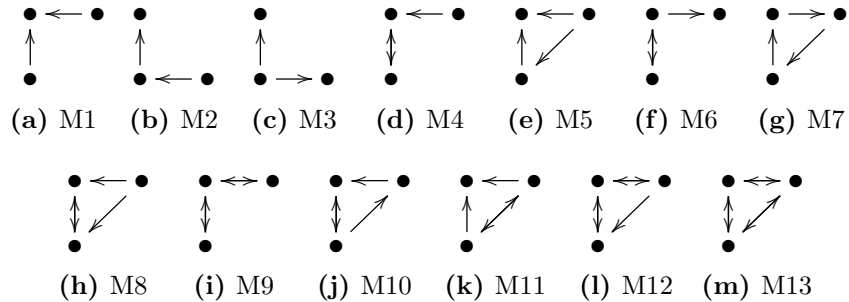
Our goal is to identify from the list presented in Fig. \ref{fig:motif} the subgraphs that are statistical significant and 
comment upon their meaning in the case of occupational mobility network ($OMN$).
For detecting them we employ \cite{Onnela1}'s alghoritm  and compute
the motif intensity $I(g)$ (\ref{eq4}) of a subgraph $g$ with links $l_g$ as the geometric mean of the weights $w_{ij}$.    
\begin{equation} 
\label{eq4}
 I(g)=(\prod_{(ij)\in l_g}{w_{ij}})^{\frac{1}{|l_g|}}
\end{equation}

where $|l_g|$ is the number of links in subgraph $g$.

The total intensity $I_M$ (\ref{eq5}) of a three-nodes motif is calculated as the sum of the its subgraph intensities $I_g$.
\begin{equation}
 \label{eq5}
I_M=\sum_{g\in M}{I(g)}
\end{equation}
There are 13 structural three-nodes motifs \ref{fig:motifs} and 
to evaluates their statistical signifance we calculate the motif intensity score \ref{eq6} for each of them.  
\begin{equation}
 \label{eq6}
\widetilde{z}_M=\frac{I_M-\langle i_M \rangle}{(\langle i_M^2 \rangle -{\langle i_M\rangle}^2)^{1/2}}
\end{equation}
where $i_M$ is the total intensity of motif M in the reference networks. The null-model network is constructed
as an ensemble of random networks generated by reshuffling the empirical weights.
\cite{Milo1}, \cite{Rubinov}.

\begin{figure}[h!]
 \centering
 \includegraphics[width=0.9\textwidth]{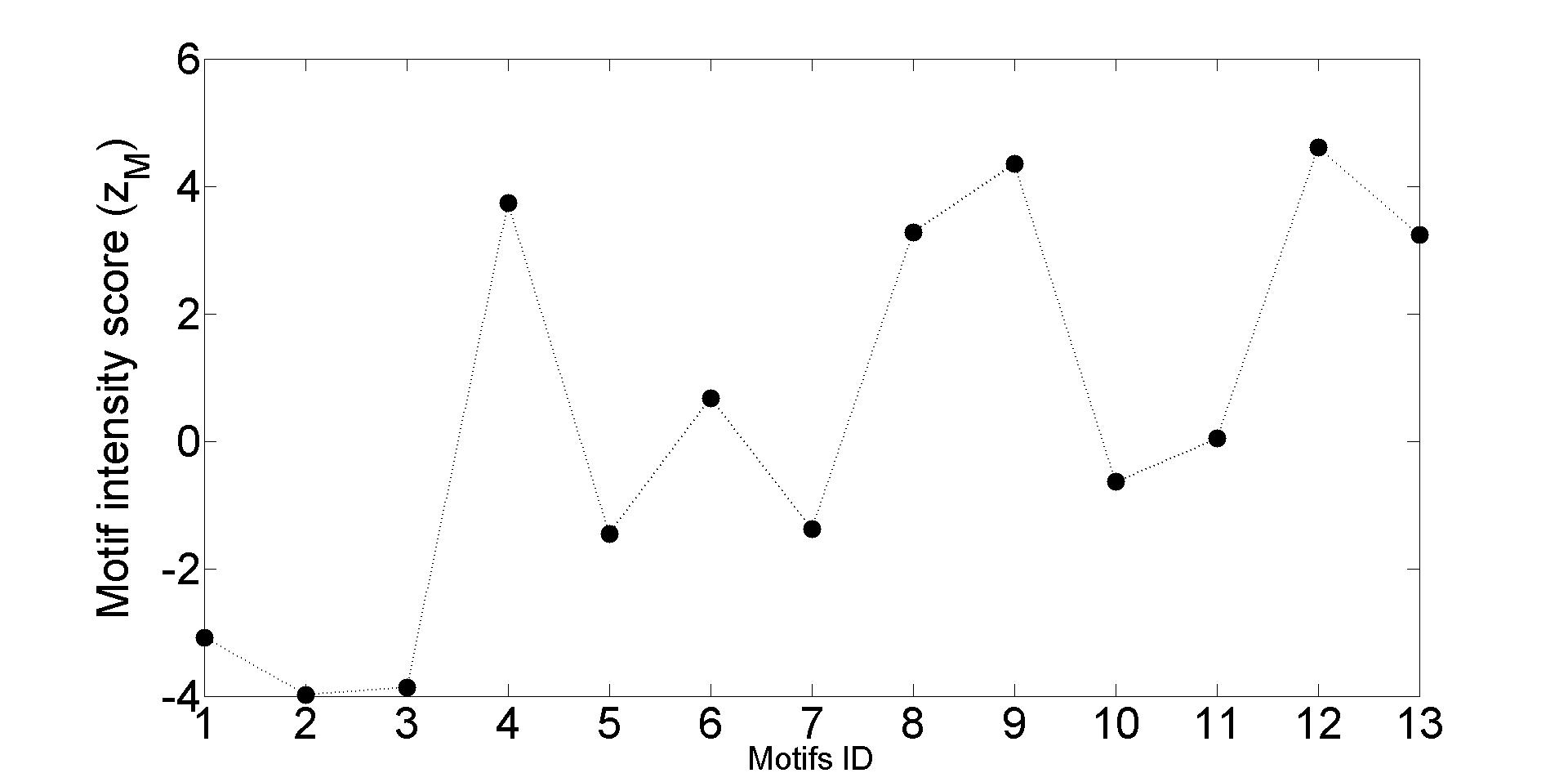}
\caption{Total motif intensities $I_M$ of the empirical network}
 \label{fig: mot}
\end{figure}

\begin{figure}[h!]
  \centering
  \subfloat[$z_{M9}=4.60$]{\label{fig:motif9}\includegraphics[width=0.33\textwidth]{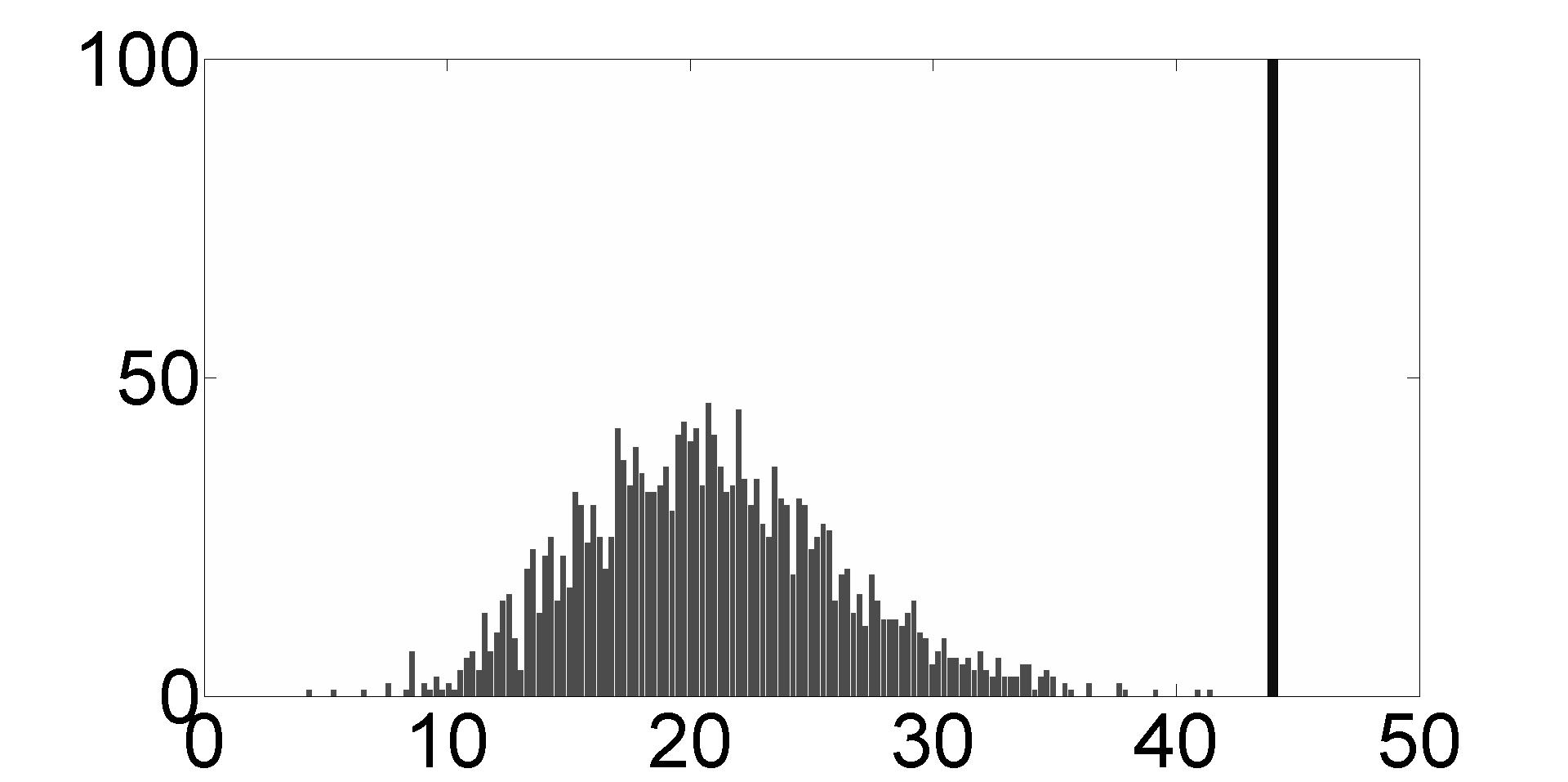}}                
  \subfloat[$z_{M12}=4.25$]{\label{fig:motif12}\includegraphics[width=0.33\textwidth]{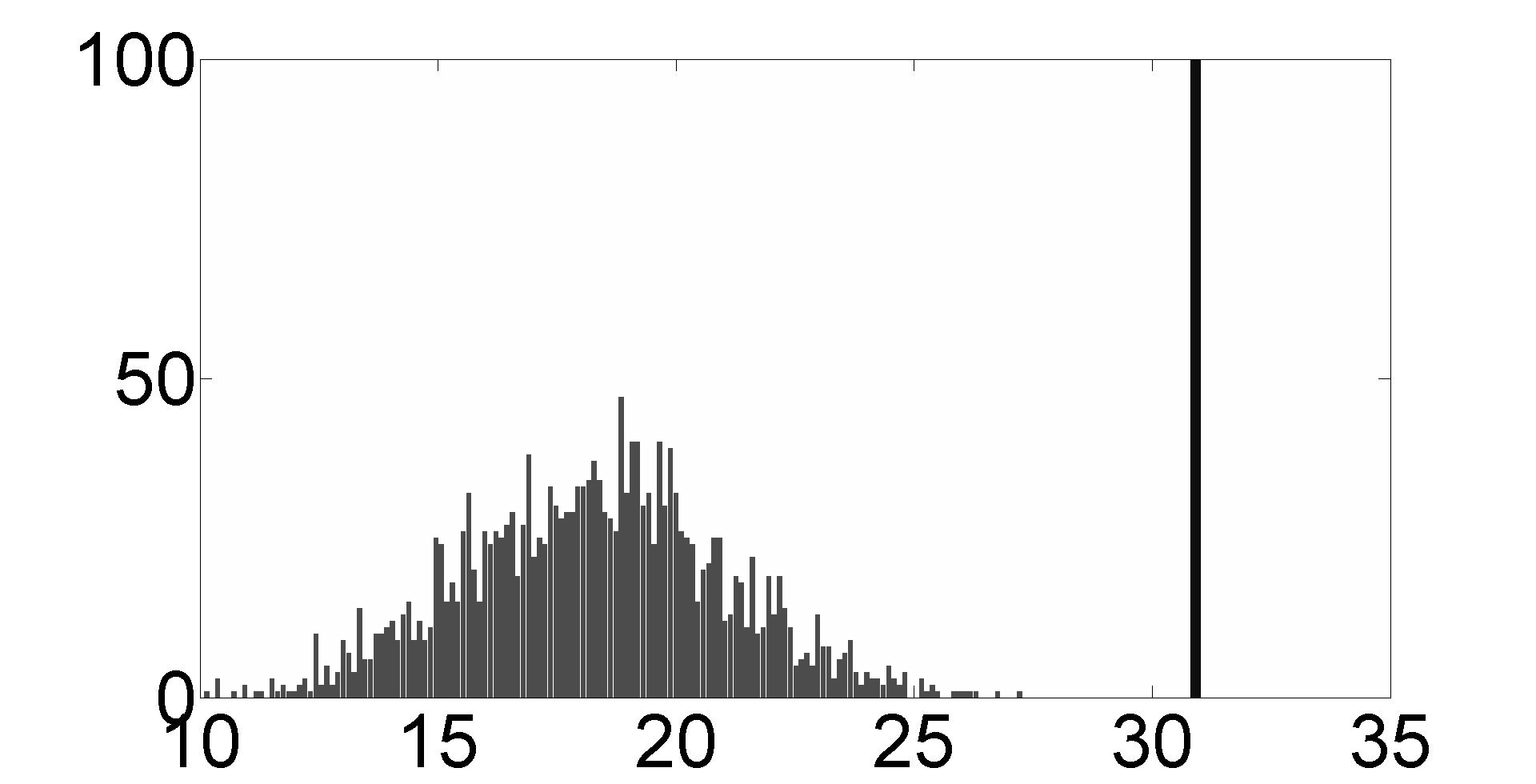}} 
  \subfloat[$z_{M4}=3.73$]{\label{fig:motif4}\includegraphics[width=0.33\textwidth]{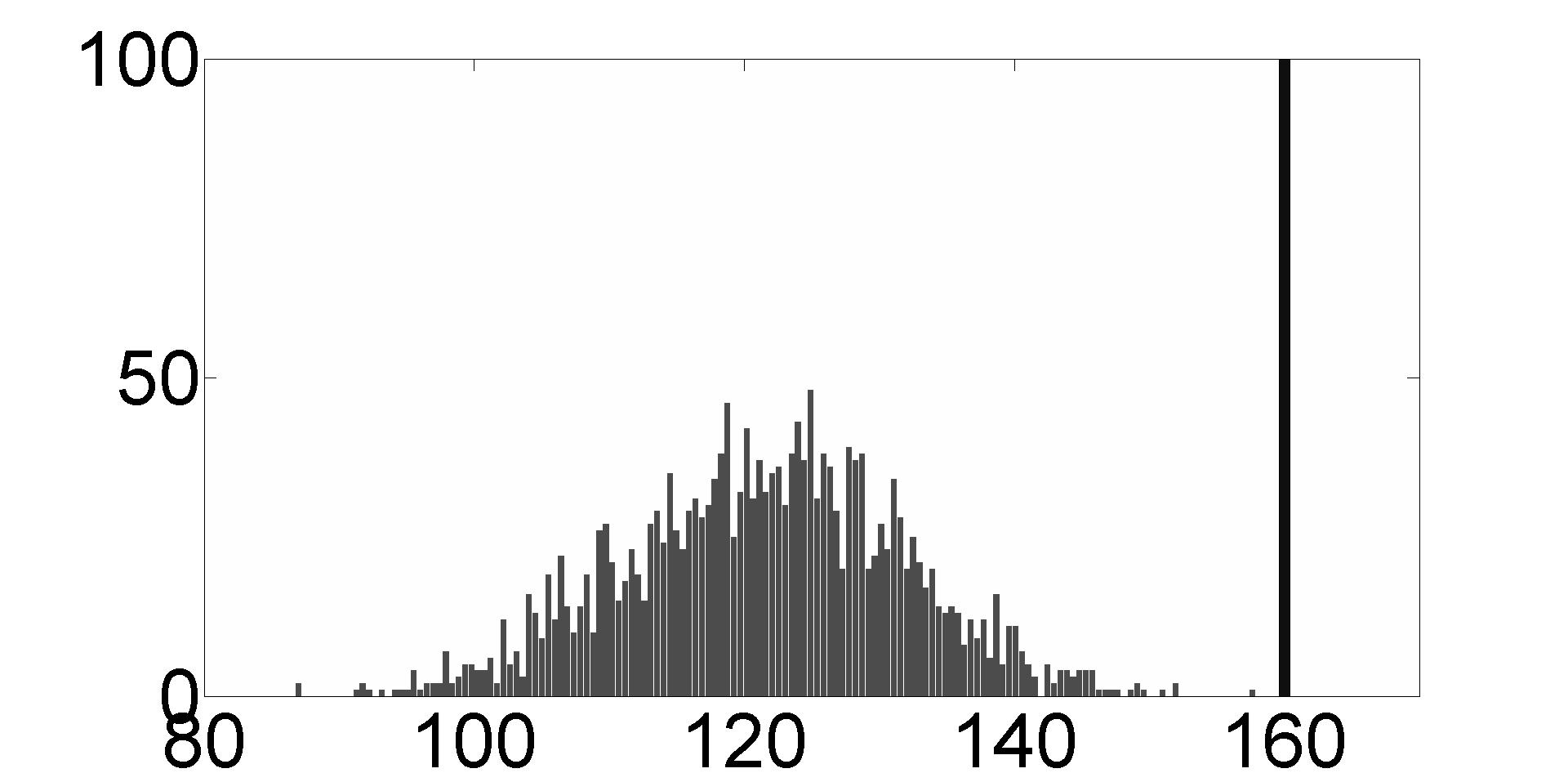}} \\
 \subfloat[$z_{M8}=3.28$]{\label{fig:motif8}\includegraphics[width=0.33\textwidth]{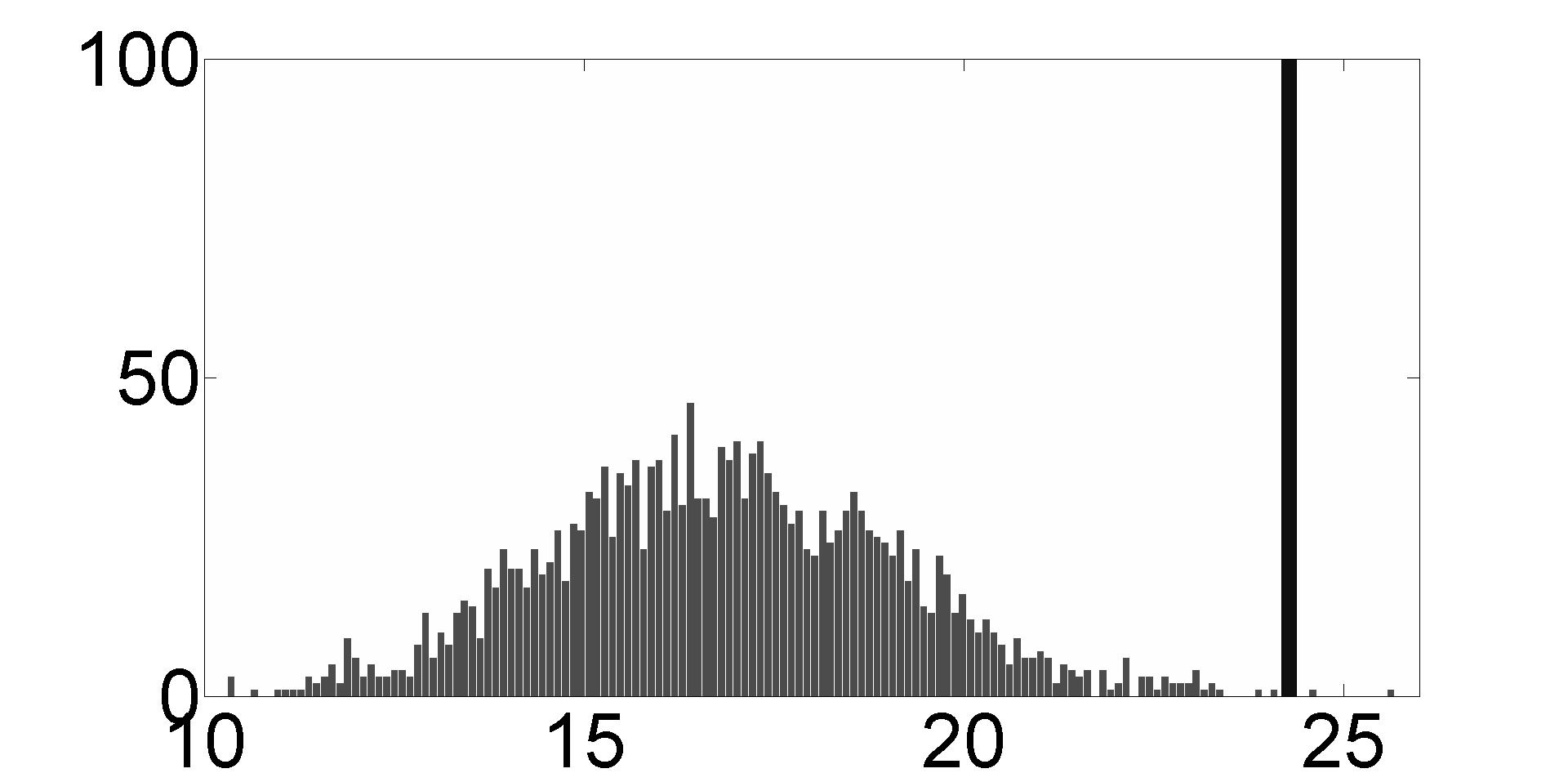}} 
 \subfloat[$z_{M13}=3.23$]{\label{fig:motif13}\includegraphics[width=0.33\textwidth]{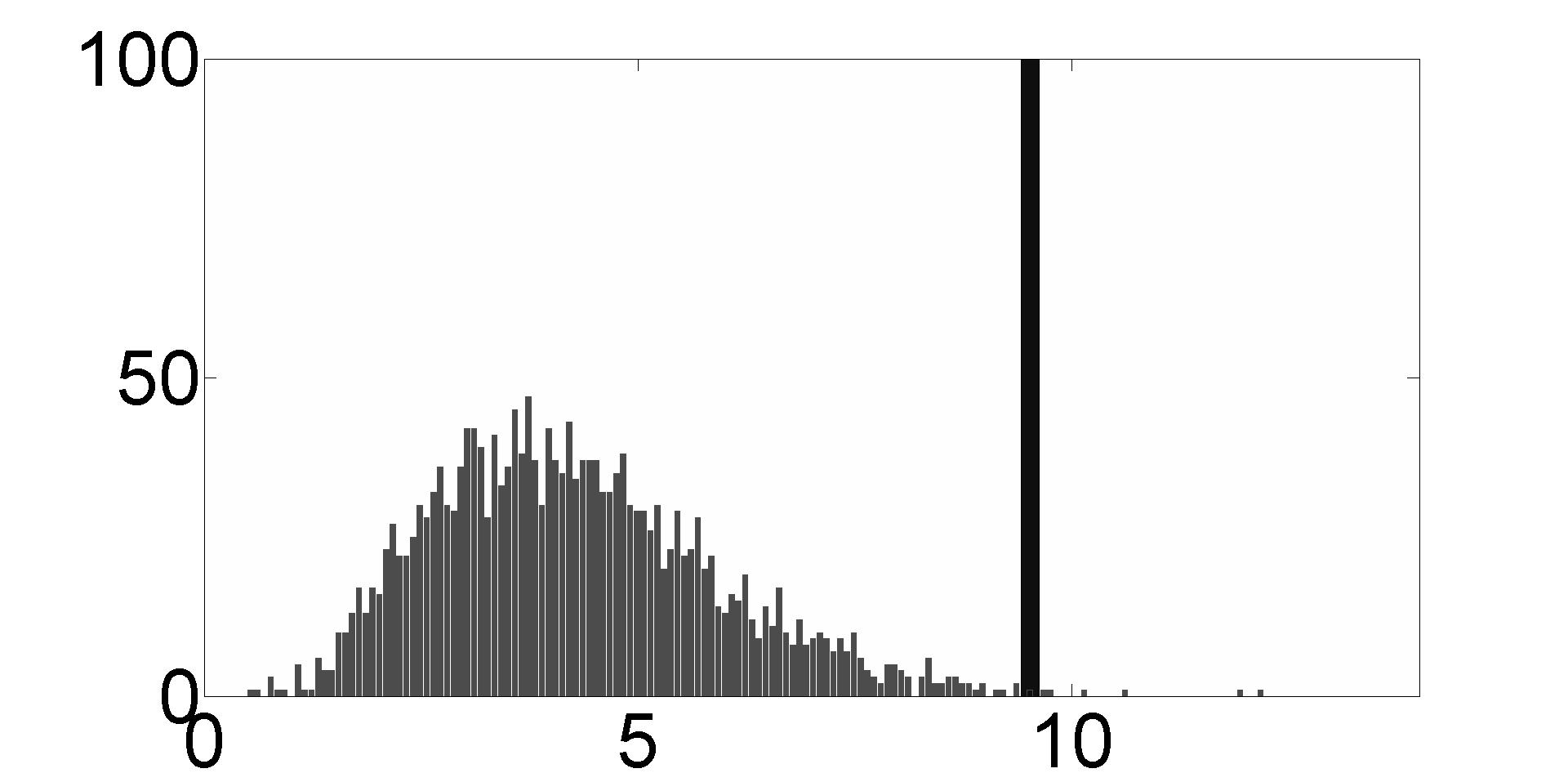}}
  \caption{Total structural motif intensities $I_M$ for the empirical network (vertical line) and the corresponding random
ensembles (histograms)}
  \label{fig:hist}
\end{figure}

Fig. \ref{fig: mot} shows the motif intensity scores for all 13 three-nodes cliques: subgraphs number 2 and 3 can be considered
structural anti-motifs and subgraphs 4, 8, 9, 12, 13 structural motifs. The total empirical intensity of the
structural motifs versus the corresponding histogram of random networks are presented in Fig. \ref{fig:hist}.

We can see that in the $OMN$ exist strong connections between certain occupations, being favorated the motifs with both
sides connections.

\begin{table}
\caption{Motif fingerprint of the first 10 occupations ordered according to their total strengh (1 - present, 0 - absent)}
\label{tab:tab1}       
\resizebox{\textwidth}{!}{
\begin{tabular}{cccccccccccccc}
\hline\noalign{\smallskip}
ISCO 88 & M1 & M2 & M3 & M4 & M5 & M6 & M7 & M8 & M9 & M10 & M11 & M12 & M13  \\
\noalign{\smallskip}\hline\noalign{\smallskip}
244 & 1 & 1 & 0 & 1 & 1 & 1 & 1 & 1 & 1 & 1 & 1 & 1 & 1 \\
214 & 0 & 0 & 0 & 1 & 0 & 1 & 0 & 1 & 1 & 0 & 1 & 1 & 1 \\
232 & 1 & 1 & 1 & 1 & 1 & 1 & 0 & 1 & 1 & 1 & 1 & 1 & 1 \\
241 & 1 & 0 & 0 & 1 & 0 & 1 & 0 & 1 & 1 & 1 & 0 & 1 & 1 \\
341 & 0 & 0 & 0 & 1 & 1 & 1 & 0 & 1 & 1 & 1 & 1 & 1 & 1 \\
122 & 0 & 0 & 0 & 1 & 0 & 1 & 0 & 1 & 1 & 1 & 0 & 1 & 1 \\
222 & 0 & 0 & 0 & 0 & 0 & 0 & 0 & 0 & 0 & 0 & 0 & 0 & 0  \\
242 & 0 & 0 & 0 & 0 & 0 & 0 & 0 & 0 & 1 & 0 & 0 & 0 & 0  \\
232 & 0 & 0 & 1 & 0 & 1 & 0 & 1 & 1 & 0 & 1 & 1 & 1 & 0  \\
213 & 0 & 0 & 0 & 0 & 0 & 0 & 0 & 0 & 1 & 0 & 0 & 1 & 0  \\
\noalign{\smallskip}\hline
\end{tabular}}
\end{table}
The frequency of occurance (motif intensity) of different motifs around a certain node can be consider as motif fingerprint.
It gives us a measure of similarity between occupation. We consider again 
the first 10 nodes ordered according to their strenght and calculate their motif fingerprint (Table \ref{tab:tab1}). A 
motif is presented around a node if it's intensity is at the right of the random network histogram and absent if the motif intensity
overlaps the histogram. 

One evidence clearly steps out here: 222 (Health professionals (except nursing))
and almost 242 (Legal professionals) don't have statistically
significant motifs around them. Around 242 is present just one type of motif (M9, Fig.\ref{fig:motifs}),
a bilateraly exchange with other two occupations, which aren't connected between them. 
These two occupations are highly specialized, require a lot of investment in
education 
and have a strict reglementation. 

\section{Conclusions}
Employing a network based approach to occupational mobility helps to identify and investigate patterns, connections between
occupations and visualise career paths. Up to now we managed to identify the central occupations in the network (node centrality)
and see what type of connections they have with the other occupations (motifs). Futher research will be directed to build this
kind of occupational mobility networks for diferent EU countries and produce a comparative analysis. Also we are interested in
explaining and the magnitude of workers flow between occupations using a gravitional model and nevertheless threat this networks
as dynamic ones.  

\section{Appendix}
\begin{longtable}{p{3 cm} p{12 cm}}

\caption{List of occupations at 3 digits according to ISCO 88 (the ones marked with (*) are present just in the Romanian
classification) \label{table:COR}}\\

\hline\noalign{\smallskip}
{\textbf{ISCO 88}} &
{\textbf{Name}}\\
\noalign{\smallskip}\hline\noalign{\smallskip}
\endfirsthead

\multicolumn{2}{c}{{\tablename} \thetable{} -- Continued} \\[0.5ex]
\hline\noalign{\smallskip}
{\textbf{ISCO 88}} &
{\textbf{Name}}\\
\noalign{\smallskip}\hline\noalign{\smallskip}
\endhead

\multicolumn{2}{l}{{Continued on Next Page\ldots}} \\
\endfoot
\noalign{\smallskip}\hline
\endlastfoot

111 & Legislators and senior government officials \\ 
114 & Senior officials of special-interest organisations \\ 
121 & Directors and chief executives \\ 
122 & Production and operations managers \\   
123 & Other specialist managers \\ 
131 & Managers of small enterprises \\  
211 & Physicists, chemists and related professionals \\ 
212 & Mathematicians, statisticians and related professionals \\ 
213 & Computing professionals \\ 
214 & Architects, engineers and related professionals \\
215* & Engineers in textile, leather and food industry \\ 
216* & Engineers in timber, glass and ceramics, pulp,paper and building materials \\  
221 & Life science professionals \\   
222 & Health professionals (except nursing) \\ 
231 & College, university and higher education teaching professionals \\ 
232 & Secondary education teaching professionals \\ 
233 & Primary and pre-primary education teaching professionals \\ 
234 & Special education teaching professionals \\ 
235 & Other teaching professionals \\ 
241 & Business professionals \\ 
242 & Legal professionals \\ 
243 & Archivists, librarians and related information professionals \\ 
244 & Social science and related professionals \\ 
245 & Writers and creative or performing artists \\ 
246 & Religious professionals \\ 
247 & Public service administrative professionals \\ 
248* & Researchers and research assistants in physical and chemical sciences \\ 
249* & Researchers and research assistants in mathematics and statistics \\ 
250* & Researchers and research assistants in informatics \\ 
251* & Researchers and research assistants in technical sciences \\ 
252* & Researchers and research assistants in the fields of textiles, leather, food industry \\ 
253* & Researchers and research assistants in wood and oxide materials \\ 
254* & Researchers and research assistants in life sciences \\ 
255* & Researchers and research assistants in medicine \\ 
256* & Researchers and research assistants banking \\ 
257* & Researchers and research assistants in legal sciences \\ 
258* & Researchers and research assistants in humanities and social sciences \\
311 & Physical and engineering science technicians \\
312 & Computer associate professionals \\
313 & Optical and electronic equipment operators \\
314 & Ship and aircraft controllers and technicians \\
315 & Safety and quality inspectors \\
316* & Technicians and foremen in textile, leather and food \\
317* & Technicians and foremen in timber, glass and ceramics, pulp, paper and building materials \\
321 & Life science technicians and related associate professional \\
322 & Health associate professionals (except nursing) \\
323 & Nursing and midwifery associate professionals \\
331 & Primary education teaching associate professionals \\
332 & Pre-primary education teaching associate professionals \\
333 & Special education teaching associate professionals \\
334 & Other teaching associate professionals \\
341 & Finance and sales associate professionals \\
342 & Business services agents and trade brokers \\
343 & Administrative associate professionals \\
344 & Customs, tax and related government associate professionals \\
345 & Police inspectors and detectives \\
346 & Social work associate professionals \\
347 & Artistic, entertainment and sports associate professionals \\
348 & Religious associate professionals \\
349* & Technicians in tourism and leisure activities \\
411 & Secretaries and keyboard-operating clerks \\
412 & Numerical clerks \\
413 & Material-recording and transport clerks \\
414 & Library, mail and related clerks \\
419 & Other office clerks \\
421 & Cashiers, tellers and related clerks \\
422 & Client information clerks \\
423* & Community officials in public service \\
511 & Travel attendants and related workers \\
512 & Housekeeping and restaurant services workers \\
513 & Personal care and related workers \\
514 & Other personal services workers \\
516 & Protective services workers \\
521 & Fashion and other models \\
522 & Shop, stall and market salespersons and demonstrators \\
611 & Market gardeners and crop growers \\
612 & Animal producers and related workers \\
613 & Crop and animal producers \\
614 & Forestry and related workers \\
615 & Fishery workers, hunters and trappers \\
711 & Miners, shotfirers, stone cutters and carvers \\
712 & Building frame and related trades workers \\
713 & Building finishers and related trades workers \\
714 & Painters, building structure cleaners and related trades workers \\
721 & Metal moulders, welders, sheet-metal workers, structural-metal preparers, and related trades workers \\
722 & Blacksmiths, tool-makers and related trades workers \\
723 & Machinery mechanics and fitters \\
724 & Electrical and electronic equipment mechanics and fitters \\
731 & Precision workers in metal and related materials \\
732 & Potters, glass-makers and related trades workers \\
733 & Handicraft workers in wood, textile, leather and related materials \\
734 & Craft printing and related trades workers \\
741 & Food processing and related trades workers \\
742 & Wood treaters, cabinet-makers and related trades workers \\
743 & Textile, garment and related trades workers \\
744 & Pelt, leather and shoemaking trades workers \\
811 & Mining and mineral-processing-plant operators \\
812 & Metal-processing plant operators \\
813 & Glass, ceramics and related plant operators \\
814 & Wood-processing- and papermaking-plant operators \\
815 & Chemical-processing-plant operators \\
816 & Power-production and related plant operators \\
817 & Industrial robot operators \\
818* & Operators of technical means of active intervention in the atmosphere \\
821 & Metal- and mineral-products machine operators \\
822 & Chemical-products machine operators \\
823 & Rubber- and plastic-products machine operators \\
824 & Wood-products machine operators \\
825 & Printing-, binding- and paper-products machine operators \\
826 & Textile-, fur- and leather-products machine operators \\
827 & Food and related products machine operators \\
828 & Assemblers \\
829 & Other machine operators not elsewhere classified \\
831 & Locomotive engine drivers and related workers \\
832 & Motor vehicle drivers \\
833 & Agricultural and other mobile plant operators \\
834 & Ships' deck crews and related workers \\
835* & Harbor workers \\
911 & Street vendors and related workers \\
912 & Shoe cleaning and other street services elementary occupations \\
913 & Domestic and related helpers, cleaners and launderers \\
914 & Building caretakers, window and related cleaners \\
915 & Messengers, porters, doorkeepers and related workers \\
916 & Garbage collectors and related labourers \\
921 & Agricultural, fishery and related labourers \\
931 & Mining and construction labourers \\
932 & Manufacturing labourers \\
933 & Transport labourers and freight handlers \\
941* & Apprentices \\

\end{longtable}

\bibliographystyle{plain}      

\bibliography{bib}   

\begin{thebibliography}{10}

\bibitem{Barabasi1}
Albert-Laszlo Barabasi, Reka Albert, and Hawoong Jeong.
\newblock {Mean-field theory for scale-free random networks}.
\newblock {\em Physica A}, 272:173--187, Jul 1999.

\bibitem{Caldarelli}
G.~Caldarelli, A.~Capocci, P.~De~Los~Rios, and M.~A. Munoz.
\newblock Scale-free networks from varying vertex intrinsic fitness.
\newblock {\em Phys. Rev. Lett.}, 89(25):258702, Dec 2002.

\bibitem{Fagiolo2}
Giorgio Fagiolo.
\newblock Clustering in complex directed networks.
\newblock {\em Phys. Rev. E}, 76(2):026107, Aug 2007.

\bibitem{Granovetter}
Mark Granovetter.
\newblock {\em {Getting a Job: A Study of Contacts and Careers}}.
\newblock University Of Chicago Press, 2 sub edition, March 1995.

\bibitem{Jackson1}
Matthew~O. Jackson and Brian~W. Rogers.
\newblock Meeting strangers and friends of friends: How random are social
  networks?
\newblock {\em American Economic Review}, 97(3):890--915, June 2007.

\bibitem{Jovanovic}
Boyan Jovanovic.
\newblock Job matching and the theory of turnover.
\newblock {\em Journal of Political Economy}, 87(5):972--90, October 1979.

\bibitem{Kambourov}
Gueorgui Kambourov and Iourii Manovskii.
\newblock Occupational specificity of human capital.
\newblock {\em International Economic Review}, 50(1):63--115, 02 2009.

\bibitem{Li}
Xiang Li, Yu~Ying Jin, and Guanrong Chen.
\newblock Complexity and synchronization of the world trade web.
\newblock {\em Physica A: Statistical Mechanics and its Applications},
  328(1-2):287 -- 296, 2003.

\bibitem{Davia}
Davia M.
\newblock Job mobility and wage mobility at the beginning of the working
  career: a comparative view across europe.
\newblock ISER working papers 2005-03, Institute for Social and Economic
  Research, January 2005.

\bibitem{McCall}
Brian~P. McCall.
\newblock Occupational matching: A test of sorts.
\newblock Working Papers 617, Princeton University, Department of Economics,
  Industrial Relations Section., Aug 1988.

\bibitem{Neal}
Derek Neal.
\newblock The complexity of job mobility among young men.
\newblock {\em Journal of Labor Economics}, 17(2):237--61, April 1999.

\bibitem{Reichardt}
J.~Reichardt and D.R. White.
\newblock Role models for complex networks.
\newblock {\em Eur. Phys. J. B}, 60(2):217--224, 2007.

\bibitem{Shaw}
Kathryn~L. Shaw.
\newblock Occupational change, employer change, and the transferability of
  skills.
\newblock Working Paper Series / Economic Activity Section~55, Board of
  Governors of the Federal Reserve System (U.S.), 1985.

\bibitem{Sicherman}
Nachum Sicherman and Oded Galor.
\newblock A theory of career mobility.
\newblock {\em Journal of Political Economy}, 98(1):169--92, February 1990.

\bibitem{Topel}
Robert~H Topel and Michael~P Ward.
\newblock Job mobility and the careers of young men.
\newblock {\em The Quarterly Journal of Economics}, 107(2):439--79, May 1992.

\bibitem{Weiss}
Yoram Weiss.
\newblock Learning by doing and occupational specialization.
\newblock {\em Journal of Economic Theory}, 3(2):189--198, June 1971.

\end{thebibliography}

\end{document}